\documentstyle[12pt]{article}

\def\thefootnote{\fnsymbol{footnote}}
\def\bea{\begin{eqnarray}}
\def\eea{\end{eqnarray}}
\def\beq{\begin{equation}}
\def\eeq{\end{equation}}
\def\tG{{\tilde G}}
\def\tF{{\tilde F}}
\def\notd{\not{\hspace{-.03in}\partial}}

\def\bs{\bar{\sigma}}
\def\[{\left [}
\def\]{\right ]}
\def\({\left (}
\def\){\right )}

\def\pp{\partial}

\def\T{\bar{T}}

\def\R{{\cal{R}}}
\def\S{{\bar{S}}}

\def\Tr{{\rm Tr}}

\def\K{{\cal K}}
\def\G{{\cal G}}
\def\f{\bar{f}}
\def\F{\bar{F}}

\def\L{{\cal L}}
\def\D{{\cal D}}

\def\tl{{\tilde\lambda}}
\def\tbl{{\bar{\tl}}}

\def\t{\bar{t}}

\def\s{\bar{s}}

\def\sr{$SL(2,\R)$}
\def\sz{$SL(2,{\cal Z})$}

\def\bp{\bar{\psi}}
\def\Z{{\bar{Z}}}

\catcode`\@=11

\begin{document}

\begin{titlepage}
\begin{center}

\hfill LBL-37198 \\
\hfill UCB-PTH-95/13 \\
\hfill LPTHE-Orsay 95/38 \\
\hfill June 1995 \\
\hfill hep-th/9506207 \\[.3in]

{\large \bf S-DUALITY CONSTRAINTS ON EFFECTIVE POTENTIALS FOR GAUGINO
CONDENSATION}\footnote{This work was supported in part by the Director,
Office of High Energy and Nuclear Physics, Division of High Energy
Physics of the U.S. Department of Energy under contract DE-AC03-76SF00098,
in part by the National Science Foundation under grant PHY-90-21139.}\\[.2in]

Pierre Bin\'etruy\\[.1in]

{\em Laboratoire de Physique Th\'eorique et Hautes Energies,\footnote{
Laboratoire associ\'e au CNRS--URA-D0063.}\\
Universit\'e Paris-Sud, F-91405 Orsay, France}\\[.1in]

{\em and}\\[.1in]

Mary K. Gaillard \\[.1in]

{\em Department of Physics and Theoretical Physics Group, Lawrence Berkeley
Laboratory,
University of California, Berkeley, California 94720}\\[.5in]

\end{center}

\begin{abstract}
We clarify the role of approximate $S$-duality in effective supergravity
theories that are the low energy limits of string theories, and show how this
partial symmetry may be used to constrain effective lagrangians for gaugino
condensation.

\end{abstract}
\end{titlepage}
\newpage
\renewcommand{\thepage}{\arabic{page}}
\setcounter{page}{1}
\def\thefootnote{\arabic{footnote}}
\setcounter{footnote}{0}

\section{Introduction}\indent
It has long been understood that abelian gauge fields can be
coupled to scalar fields in such a way that the equations of motion are
invariant under a group of ``duality'' transformations~\cite{mkbz}
that interchange the gauge field strength
$F_{\mu\nu}$ with its dual $\tF_{\mu\nu}$, provided the (noncanonical)
kinetic energy term of the scalar field(s) [hereafter referred to as the
dilaton(s)] that couple to the gauge fields satisfies certain constraints.
This is
an automatic feature of ungauged extended supergravity theories~\cite{ext}.
A special example of this class of models is the dilaton plus Yang-Mills
sector of effective $N=1$ supergravity theories obtained from superstrings in
the limit where the
gauge and Yukawa couplings are set to zero.  In this case the group of duality
transformations is \sr, which has as a discrete subgroup \sz ~that
includes the transformation $S\to 1/S$, often referred to as
$S$-duality.  This is analogous to the modular group, known to be
exact to all orders of string perturbation theory~\cite{mod},  which
contains a subgroup \sz ~that includes the transformation
$T\to 1/T$, where $T$ is a modulus chiral supermultiplet. The $vev$ of its
scalar component $t$  determines the radius of compactification.
In models from orbifold compactification,
this discrete symmetry is generally a subgroup of a classical continuous
symmetry that contains \sr ~as a subgroup. Such symmetries are anomalous at
the quantum level of the effective field theory since, for example,
in supergravity they entail chiral transformations on fermions.  In the case of
modular invariance counterterms~\cite{dkl,counter} must be added to the
effective field theory so as to restore the discrete modular symmetry.

It has been conjectured~\cite{iban} that a similar situation might hold with
respect to $S$-duality.  Since the $vev$ of the complex scalar field $s$ that
is the scalar member of the dilaton supermultiplet $S$ determines the gauge
coupling constant: $\langle s\rangle = g^{-2} - i\theta/8\pi^2$, this
corresponds to
strong/weak coupling duality.  Recently there has been considerable interest
in $S$-duality from both the string~\cite{john,wit} and
field~\cite{john,seib,lalak}
theory points of view. In particular, it has been shown that the \sz~
subgroup of \sr~ that is generated by the elements  $\theta\to\theta+2\pi$ and
$4\pi s\to 1/4\pi s$ relates different string theories~\cite{wit}, and also
that certain theories are $S$-duality invariant~\cite{seib} under
transformations involving both elementary fields and nonperturbative solutions.

In this paper we restrict our attention to effective field theories of the type
that have been obtained explicitly in orbifold compactifications.  In these
theories $S$-duality, which we define hereafter as the \sr ~group of duality
transformations among elementary fields, is a symmetry of the equations of
motion only of the dilaton-gauge-gravity sector in
the limit of vanishing gauge couplings -- that is, the limit in which the gauge
group reduces to $U(1)^n$, where $n$ is the total number of gauge degrees of
freedom.  Since the effective superpotential~\cite{ven,tom} that parameterizes
gaugino condensation is induced by the nonabelian self-couplings of Yang-Mills
fields and by their gauge couplings to chiral supermultiplets, this term must
explicitly break duality.  Nevertheless, approximate $S$-duality may be a
useful tool in parameterizing additional quantum effects that arise from the
couplings of the Yang-Mills sector to the gravity and dilaton sectors.
In Section 2 we show how the dilaton couplings in the chiral multiplet
formulation are related to the general formulation~\cite{mkbz} of duality
invariance for interacting vector fields, and use this formulation to obtain
the
duality transformation law for the composite chiral multiplet that is
interpreted as the lightest bound state of a confined Yang-Mills sector.
In section 3 we give examples of how this transformation property may be used
to constrain effective potentials in the chiral formulation.  We will consider
both the chiral multiplet and the linear multiplet formulations for the
dilaton superfield.

\section{Duality transformations}\setcounter{equation}{0}\indent
We first recall the relevant elements of the general formalism constructed
in~\cite{mkbz} for a noncompact group $\G\subset Sp(2n,\R)$ of duality
transformations on $n$ vector field strengths $F$.  The scalars are valued
in the coset space $\G/\K$, where $\K\subset U(n)$ is the maximal
compact subgroup of $\G$, and they can be represented by a group element $g$:
\beq g = \pmatrix{\phi_0&\phi_1^*\cr\phi_1&\phi_0^*\cr} ,
\;\;\;\;  \phi_0^{\dag}\phi_0 - \phi_1^{\dag}\phi_1 = 1,\eeq
that is restricted to the special form:
\beq g = \exp\pmatrix{0&P^*\cr P&0\cr}.\eeq
Under $\G$ the vector field strengths and scalar fields, respectively,
transform as
\bea \pmatrix{F+iG\cr F-iG\cr} &\to& g_0\pmatrix{F+iG\cr F-iG\cr},
\;\;\;\; g_0= \exp\pmatrix{T&V^*\cr V&T^*\cr}, \nonumber\\
g &\to& g_0gk^{-1} = \exp\pmatrix{0&P'^*\cr P'&0\cr},\eea
where $\tG^{\mu\nu} = {1\over 2}\epsilon^{\mu\nu\rho\sigma}G_{\rho\sigma}
= 2\pp\L/\pp F^{\mu\nu}$,
$T = -T^{\dag}$ is a compact generator of $\K$, $V = V^T$ is a
noncompact generator of $\G/\K$, and $k= k(g_0,P)$ is an element
of $\K$ with field-dependent parameters chosen so as to preserve the
off-diagonal form of ln$g$. The symmetry of $P$ implies $\phi_0^{\dag} =
\phi_0,\; \phi_1^T = \phi_1.$  The lagrangian describing these bose degrees of
freedom, is\footnote{The normalization of the vector kinetic term here differs
by a factor two from that in~\cite{mkbz}, where it was chosen to coincide with
the canonical one in the limit $Z\to 0$.}
\bea \L &=& -{1\over 4}F_{\mu\nu}{\rm Re}fF^{\mu\nu} -{1\over 4}F_{\mu\nu}
{\rm Im}f\tF^{\mu\nu} + {1\over2}\Tr\(P^\mu P_\mu\),\qquad
f = {1\over2}\({1-Z^*\over 1 + Z^*}\),\nonumber\\
Z &=& Z^T = \phi_1(\phi_0)^{-1}, \qquad
P_\mu = g^{-1}D_\mu g \equiv g^{-1}\pp_\mu g - Q_\mu, \eea
where $Q_\mu$ and $P_\mu$ are the parts in the Lie algebra of $\K$ and of
$\G/\K$, respectively, of the the element $g^{-1}\pp_\mu g = Q_\mu + P_\mu$
of the Lie algebra of $\G$.  The equations of motion derived from the
lagrangian (2.4) are invariant under the transformations (2.3), although the
action is not.\footnote{An invariant action for supergravity with an
abelian gauge group has been constructed in~\cite{john} by sacrificing
manifest general coordinate invariance.}
Couplings to other fields $\psi_i$ are determined by their transformation
properties under $\K$; the equations of motion of the following lagrangian
are invariant under (2.3) with $\psi_i\to k_i(g_0,P)\psi_i$,
where $k_i$ represents $\K$ on the multiplet of fields $\psi_i$:
\beq\L(\psi)= \bp\gamma^\mu D_\mu\psi + \alpha^{ij}\bp_i\gamma^\mu P_\mu\psi_j
+ \cdots, \;\;\;\; D_\mu\psi = \pp_\mu + Q_\mu\psi, \eeq
where the $\alpha^{ij}$ are constants, and
the dots represent possible $\G$-invariant operators of higher dimension, as
well as operators constructed from the field strength $F_{\mu\nu}$, an
arbitrary
antisymmetric tensor $H_{\mu\nu}(\psi)$, and their duals~\cite{mkbz}.

Here we are interested in the simplest realization of the above construction,
namely $\G=$ \sr, $\K = U(1).$   Then the $n\times n$ matrices $\phi_i,
T,V,Z,P$ and $f$ are proportional to the unit matrix.  The dilaton $s$ from
superstring theory is identified as
\beq s = {1\over2}\({1-Z^*\over 1+Z^*}\) = f.\eeq
Expressing $\phi_i$ in terms of $s$ we obtain:
\bea  P_\mu &=& -{1\over s+\s}\pmatrix{0&(\eta^*)^2\pp_\mu\s\cr
\eta^2\pp_\mu s&0\cr},\qquad {1\over2}\Tr(P_\mu P^\mu) = {\pp_\mu s\pp^\mu\s
\over s+\s},\nonumber \\ \eta &=& (\eta^*)^{-1} = \sqrt{{1+2\s\over1+2s}}. \eea
Inserting (2.6) and (2.7) in (2.4), we recover the standard lagrangian for the
string dilaton with K\"ahler potential $K(S,\S) = -\ln(S+\S).$
In addition we have
\beq  Q_\mu = {1\over2(s+\s)}\pmatrix{\pp_\mu s - \pp_\mu\s&0\cr 0&
\pp_\mu\s - \pp_\mu s\cr} + \pmatrix{\eta^*\pp_\mu\eta &0 \cr
0&\eta\pp_\mu \eta^*\cr}. \eeq
The action of \sr ~on $s$ takes the usual form
\beq s\to s' = {as -ib\over ics + d},\qquad a,b,c,d\in\R,\qquad ad-bc =1.
\eeq  This corresponds in (2.3) to
\bea g_0 &=& \pmatrix{p&q^*\cr q&p^*\cr},\qquad k
= \pmatrix{\zeta&0\cr0&\zeta^*\cr},
\nonumber \\ p &=& {1\over2}(a+d) + {1\over4}c -ib,\qquad q
= {1\over2}(d-a) + {1\over4}c +ib, \nonumber \\
\zeta &=& (\zeta^*)^{-1} = {\eta\over\eta'}\xi^*, \qquad\eta'=\eta(s'),
\qquad \xi =\sqrt{-ic\s+d\over ics+d}.\eea
A Weyl fermion $\psi_n$ of $U(1)$ charge $n$ transforms as
\beq \psi_n \to \psi'_n = \zeta^n\psi_n,\eeq and its \sr ~invariant kinetic
energy is given by
\bea &&\L_{K.E.}(\psi_n) = \bp_n\gamma^\mu D_\mu\psi_n = \bp_n\gamma^\mu
\(\pp_\mu + {in\over s+\s}\pp_\mu{\rm Im}s + n\eta^n\pp_\mu \eta\)\psi_n
\nonumber \\ &&\qquad = {i\over(s+\s)^{2\beta}}\f_{n,\beta}\gamma^\mu
\(\pp_\mu - {2\beta\over s+\s}\pp_\mu{\rm Re}s + {in\over s+\s}\pp_\mu
{\rm Im}s\)f_{n,\beta}, \eea
where we have defined $f_{n,\beta} = \eta^n(s+\s)^\beta\psi_n$; under \sr:
\beq f_{n,\beta}\to f'_{n,\beta} = \xi^{-n}|isc + d|^{-2\beta}
f_{n,\beta} = \xi^{-2\beta-n}(ics + d)^{-2\beta}f_{n,\beta}.\eeq

In a supersymmetric theory the transformation property (2.9) of $s$ applies to
the dilaton chiral supermultiplet $S(\theta) = s + \theta\chi_S + \cdots$:
\beq S(\theta)\to S'(\theta) = {aS(\theta') - ib\over icS(\theta') + d}.\eeq
This effects a K\"ahler transformation:
\beq K\to K + F + \F, \qquad F = \ln(ics+d), \qquad \theta' =
e^{{1\over4}(F-\F)
}\theta = \xi^{-{1\over2}}\theta, \eeq
from which one can derive the transformation properties of the component fields
of $S$; in particular
\beq \chi_S\to\chi'_S = \xi^{-{1\over2}}(ics + d)^{-2}\chi_S. \eeq
Thus we can identify $\chi_S\equiv f_{-{3\over2},1}$ in (2.13).  It is
straightforward to verify that $\L(f_{-{3\over2},1})$ given in (2.12) coincides
with the kinetic energy term for $\chi_S$ in the standard supergravity
lagrangian~\cite{crem,bggm}.

The other fermion fields that couple to the dilaton are the gauginos $\lambda$
with kinetic energy term: \beq
\L_{K.E.}(\lambda) = {1\over2}\tbl_L\(\notd + {i\over2}{\notd{\rm Im}s\over
s+\s}\)\tl_L + {\rm h.c.}, \qquad \tl = \sqrt{{\rm Re}s}\lambda. \eeq
Comparing (2.17) with (2.12), the
(canonically normalized) field $\tl_L$ is identified as $\tl_L\equiv
f_{{1\over2},0}$, from which it follows that $\lambda_L\equiv
f_{{1\over2},-{1\over2}}$, with its transformation law given by (2.13) as:
\beq \lambda_L \to \lambda'_L = \xi^{1\over2}(ics+d)\lambda_L . \eeq
This immediately gives the transformation law for the gauge field strength
supermultiplet $W_\alpha$, where $\alpha$ is a Dirac index:
\beq  W_\alpha(\theta)\to W'_\alpha(\theta) = \xi^{1\over2}(ics+d)W_\alpha
(\theta').\eeq

Having established the transformation laws for the superfields $S$ and
$W_\alpha$, and using the invariance under \sr ~of the supervielbein, one can
directly check that the supergravity equations of motion~\cite{bggm} are
invariant under \sr ~when the the theory is limited to this set of fields.  In
this case they reduce to:
\bea
&& R = R^{\dag} = 0, \nonumber \\ &&
G_b + {1\over8(s+\s)^2}\bs^{\dot\alpha \alpha}_b \D_\alpha S\D_{\dot\alpha}\S
- {(S+\S)\over8}\bs^{\dot\alpha \alpha}_b W_\alpha W_{\dot\alpha} = 0,
\nonumber \\ &&
{S\over 2}\D^\alpha W_\alpha - {1\over2}\D_\alpha S W^\alpha + {\rm h.c.}
= 0, \nonumber \\ &&
(S +\S)^{-2}\F^S + {1\over4}W^\alpha W_\alpha = 0,\eea
where $b$ is a Lorentz index, $F^S = -{1\over4}\D^\alpha\D_\alpha S$,
and the superfields $R$ and $G_b$ describe the supergravity multiplet; they are
related to the curvature superfield by:
$$ R^{\dot\delta\dot\gamma}_{\;\;\;\;ba} =
8\(\bar{\sigma}_{ba}\epsilon\)^{\dot\delta\dot\gamma}R, \;\;\;\;
R_{\delta\;\;ba}^{\dot\gamma} =
-2iG^d\(\sigma^c\epsilon\)_\delta^{\dot\gamma}\epsilon_{dcba}.$$
Using the Bianchi
identity:
\beq \D^\alpha W_\alpha = \D_{\dot\alpha}W^{\dot\alpha},\eeq it is
straightforward to show that these equations are invariant under the
\sr ~transformations defined by (2.14) and (2.19).  It follows from the latter
that the composite superfield
\beq U\sim W^\alpha W_\alpha\eeq used in effective
supergravity lagrangians for gaugino condensation transforms as
\beq U(\theta)\to U'(\theta) = \xi(ics+d)^2U(\theta).\eeq
As shown in~\cite{us}, since the K\"ahler weight of the chiral superfield
$W_\alpha$ differs from that of an ordinary chiral superfield such as $S$, the
composite field $U$, in the context of supergravity, is related to an ordinary
chiral field $H$ by:
\beq U = e^{K/2}H^3, \eeq
where $K$ is the full K\"ahler potential; classically:
\beq K = -\ln(S+\S)+ G(Z,\Z),\qquad U= {e^{G/2}H^3\over(S+\S)^{1\over2}},\eeq
where $Z$ represents chiral multiplets other than the dilaton.
Since  under (2.14) $(S+\S) \to |icS + d|^{-2}(S+\S)$, we obtain the
transformation property
\beq H(\theta)\to H'(\theta) = (icS + d)^{1\over 3}H(\theta).\eeq

We emphasize that, although the equations of motions (2.20) make no explicit
reference to the nonabelian nature of the gauge field strength, they are
duality
invariant\footnote{There is evidence~\cite{seib} for invariance under
duality transformations involving soliton solutions in $N=2$ supersymmetric
Yang-Mills theories with vanishing $\beta$-functions (which includes the case
$N=4$).} only in the abelian case~\cite{deser}. This is because the covariant
spinorial derivatives (or, in the case of nonsupersymmetric theories, the
covariant derivatives) have an implicit dependence on the gauge potential.
Furthermore, even in the abelian case duality is broken if
gauge couplings to matter are introduced, since this also involves the gauge
potential explicitly.  Finally, $S$-duality is broken by the presence of a
superpotential $W(Z)$ through the appearance of the noninvariant factor
$e^{K(s,\s)}$ in the corresponding scalar potential and Yukawa couplings.

Finally we note that while the equations of motion (2.20) are \sr~ invariant,
the Yang-Mills superfield lagrangian~\cite{bggm} is not\footnote{See
however~\cite{john}.}:
\beq \L_{YM} = {1\over8}\int d^4\theta {E\over R}SW^\alpha W_\alpha + {\rm h.
c.}, \quad SW^\alpha W^\alpha \to |icS+d|\(a - {ib\over S}\)S W^\alpha
W^\alpha.
\eeq  This result will have implications for attempts to impose approximate
\sr \newline invariance on effective lagrangians for gaugino condensation, to
be
discussed in the next section.

\section{Gaugino condensation}\setcounter{equation}{0}\indent
\subsection{Chiral multiplet formulation}
The superpotential for gaugino condensation was first derived by Veneziano and
Yankielowicz~\cite{ven} in the context of a supersymmetric renormalizable
Yang-Mills theory, by imposing the correct chiral and conformal anomalies.
Their result was extended to include the dilaton by Taylor~\cite{tom}.
In the superfield formulation~\cite{bggm} of supergravity this leads to
the potential term~\cite{us}, expressed in terms of the chiral multiplet $H$
introduced in (2.24), of the form
\bea \L_{pot} &=& \L_C + \L_Q =
{1\over2}\int d^4\theta {E\over R}e^{K/2}\(W_C + W_Q\),
\nonumber\\ W_C &=& {1\over4}SH^3, \qquad W_Q = {b_0\over2}H^3\ln(H/\mu), \eea
with $b_0$ the group theory constant that determines the appropriate
$\beta$-function. $\L_C$ and $\L_Q$ are usually
interpreted as the classical and quantum contributions, respectively, to the
effective potential for gaugino condensation.  As anticipated in the
introduction, $\L_Q$ is not \sr~ invariant.  However, neither is $\L_C$,
since, by
construction, it has the same transformation property as (2.27). In the
general formulation~\cite{mkbz} of duality transformations, couplings of the
dilaton(s) to matter entail duality invariance of the corresponding terms
in the {\it lagrangian}, as opposed to their couplings to gauge fields, which
are only
an invariance of the {\it equations of motion.}   This mismatch may be traced
to
the fact that we used the Bianchi identity (2.21) to obtain the invariance of
the equations of motion for the underlying theory expressed in terms of the
Yang-Mills field strength $W_\alpha$.  The identification (2.22), together with
the constraint
\beq \(\D^\alpha\D_\alpha - 24R^{\dag}\)W^\beta W_\beta -
\(\D_{\dot\alpha}\D^{\dot\alpha}- 24R\)
W_{\dot\beta}W^{\dot\beta} = {\rm total \; derivative}, \eeq
implies a constraint on the superfield $H$ that is not
satisfied for an ordinary superfield.  This suggests a possible inconsistency
in
all chiral multiplet formulations of gaugino
condensation~\cite{ven,tom,us,eta,us2}, especially those
treatments~\cite{ross} \`a la
Nambu-Jona-Lasinio in which the use of a Lagrange multiplier imposes the
operator identity $U = W^\alpha W_\alpha$.  We will return to this point in
section 3.B below.

In this paper we consider a toy model with a single modulus
superfield $T$, and set gauge nonsinglet matter fields to zero.  These
simplifications do not affect the generality of our results.  In~\cite{us} we
showed that the (continuous) modular invariance of the classical supergravity
theories requires that the K\"ahler potential depend on the composite field
$H$ only through the invariant $|H|^2/(T + \T)$, and we adopted the no-scale
form
\bea  K &=& -\ln(S + \S) -3\ln(T + \T) - 3\ln\(1-f(S,\S){|H|^2\over T + \T}\)
\nonumber \\ &=& -\ln(S + \S) -3\ln\(T + \T -f(S,\S)|H|^2\).\eea
The effective theory constructed by combining (3.3) with (3.1) was
subsequently modified~\cite{eta,us2} by including an additional $T$-dependence
through the Dedekind function $\eta(T)$ in such a way as to restore invariance
under the discrete modular
group.  Although this modification was regarded as a parameterization of
threshold corrections~\cite{dkl} that arise from integrating out heavy string
modes in some orbifold compactifications, none of the above models is truly
consistent with string theory, since they do not incorporate the Green-Schwarz
counterterm~\cite{counter} that cancels part (or in many cases all~\cite{ant})
of the modular anomaly that arises from perturbative field theory quantum
corrections. The effect of the Green-Schwarz term is to modify the K\"ahler
potential by
\beq\ln(S + \S)\to \ln(S+\S - bG),\qquad G = -3\ln(T+\T),\eeq
where $b = 2b_0/3$, and $b_0 = .56$ is the constant that determines the
$E_8 \;\beta$-function.
This modification spoils the no-scale feature of the K\"ahler potential and
generally leads to an unbounded potential.

In this section we consider a prototype model in which the hidden gauge
group is $E_8$, in which case the anomaly is completely cancelled by the
Green-Schwarz term, and furthermore there is no ambiguity in constructing an
effective composite potential.  That is, since the modular transformation laws
are now, with $\alpha,\beta,\gamma,\delta\ni \R,\; \alpha\delta -\beta\gamma
=1$:
\bea T&\to&T' = {\alpha T - i\beta\over i\gamma T + \delta},\;\;\;\;
G\to G + F + \F, \qquad F = 3\ln(i\gamma T+\delta)\nonumber \\
H&\to&H' = e^{-F/3}H,\;\;\;\; S\to S'= S + bF,\eea
the structure (3.1), which can be understood~\cite{np} as arising from
coupling constant renormalization,
coincides with the requirements of modular invariance.

We study the effective lagrangian for the composite superfield $H$ under
the assumptions that modular invariance is exact to all orders, and that
$S$-duality, as defined by (2.14) and (2.26), with $T\to T'=T$, is recovered to
leading order in $g^2 \equiv $ (Re$s)^{-1}\equiv\sigma$.
$S$-duality determines the function $f(S,\S)$ in (3.3):
$f(S,\S) = (S + \S)^{1\over 3}$; we therefore adopt the K\"ahler potential:
\bea K &=& -\ln M - 3\ln(1 - (M')^{1\over 3}Q) + G, \;\;\;\; L^{-1} = S+\S -bG
, \nonumber \\ M &=& L^{-1} + 3b_1\ln Q, \;\;\;\; M' = L^{-1} + 3b_2\ln
Q,\;\;\;
\; Q = |H|^2e^{G/3}.\eea
The rationale is as follows. Modular invariance requires $K = G + k(L, Q)$.
In the limit $g^2\to 0,\;i.e.,\;\sigma\to\infty$, (3.6) is $S$-duality
invariant and reduces to the form (3.3).  By matching string one-loop
calculations with field theory
ones, it was shown in~\cite{tommk} that at the string scale the K\"ahler
potential is precisely (3.6) with $b_1=b_2=H=0$; the modular invariant scalar
field\footnote{We use reduced Planck mass units: $8\pi G_N=1/m_{Pl}^2=1$, and
throughout this subsection we
use upper case letters for chiral scalar superfields
and lower case for the corresponding complex scalars.}
\bea \ell = [s + \s + 3b\ln(t+\t)]^{-1} = 2g^2\eea
is twice the squared gauge coupling constant at
that scale~\cite{tommk,jan}.  However, we are interested in the effective
theory
at the condensation scale; one possibility is that we should replace everywhere
the string scale coupling constant by the running coupling constant evaluated
at
the condensation scale, which would correspond to $b_1 = b_2 = b.$  In this
case
the theory is of the no-scale form, and the potential is positive
semi-definite.

For the superpotential we take the most general form consistent with modular
invariance of string perturbation theory:
\beq W = c + W_0, \;\;\;\; W_0 = e^{-S/b}F(Y),\;\;\;\; Y = He^{S/3b} \eeq
that is, we take $W_0(S,H)$ to be modular invariant but allow for a constant
term that breaks modular invariance, that might arise~\cite{dine} from a
classical nonperturbative effect such as the $vev$ of the three-form
$H_{\ell mn}$ of ten dimensional supergravity.  In the standard
formulation~\cite{ven,tom,us,eta,np,us2} of gaugino condensation $F(Y)\propto
Y^3\ln Y$.
However, as noted above this is not invariant under $S$-duality in the limit
$g^2\to 0\;(\sigma\to\infty)$. If we adopt the point of view that $S$-duality
should be recovered in this limit, we require
\beq \lim_{y\to\infty}F(y)\sim y^n(\ln y)^m,\qquad n< 3. \eeq
Such a superpotential could be interpreted as arising from purely
nonperturbative effects, with, for $b_2=b$, the Yang-Mills wave function
renormalization encoded in the K\"ahler potential, rather than the
superpotential -- in contrast to (3.1).  A similar reinterpretation was
used in~\cite{us2} to recover a positive semi-definite no-scale potential in
the
presence of $\eta(T)$-dependent terms.

We have studied the potential for three choices of $b_1,b_2$ in (3.6) that we
enumerate below.  An interesting question that we address is whether
or not it is possible to generate a bounded potential (more precisely, one with
vanishing vacuum energy) with supersymmetry breaking without the introduction
of
a constant $c$ in the superpotential.  The K\"ahler potentials we consider are
by no means the most general.  For example, one-loop corrections~\cite{kamran}
induce a term:
\beq \L + \L_{1-{\rm loop}} \ni - 3\int d^4\theta E
+ {N_G\over128\pi^2}\int d^4\theta E(S+\S)^2
|W^\alpha W_\alpha|^2\ln\Lambda^2,\eeq
where $\Lambda$ is the effective cut-off, that we take here to be a constant,
and $N_G$ is the number of gauge degrees of freedom.
Since this term arises only from
couplings of the gauge sector to the dilaton sector it is $S$-duality
invariant, and would by itself generate
a kinetic term for the composite field $H$; making the replacement
$W^\alpha W_\alpha\to U$, the classical K\"ahler potential is modified as
\bea e^{-K/3}&\to&e^{-K/3}\(1 - {\alpha\over3} e^K(S+\S)^2|H|^6\),\qquad
\alpha = {N_G\ln\Lambda^2\over128\pi^2}, \nonumber \\
K&\to& K - 3\ln\(1 - {\alpha\over3}{S+\S\over(T+\T)^3}|H|^6\) + O(\hbar^2),\eea
where the $O(\hbar^2)$ terms include the substitution $S+\S\to L^{-1}$ when
exact modular invariance is imposed.  Similar higher dimensional operators
arise~\cite{bgt} from string corrections even at the classical level; it
remains to be
seen whether or not such a K\"ahler potential allows for a viable effective
potential for the composite multiplet.  Since the models we consider possess a
continuous modular symmetry, the $vev\;\langle t\rangle$, that fixes the radius
of compactification,
is undetermined.  This degeneracy of the vacuum could be lifted
by quantum corrections in the effective field theory and/or by string
corrections such as $\eta(T)$-dependent threshold corrections.

\subsubsection{$b_1 = b_2 = 0$.}  This corresponds to using only the string
coupling constant in the K\"ahler potential.  We considered a parameterization
of the superpotential $F(y)=\sum_na_ny^n$.
The requirement that the potential be positive semi-definite constrains the
values of $n$: $n> .4238$ or $n< - 4.74$ (and thus $c=0$).
We studied the potential as a function of $\ell$
for $F(y) = y^n,$ for various values of $.424\le n\le 2$ and of $x =
(m')^{1\over3}q$ with $0\le x <1$, as required by positivity of the
scalar metric. We found that, if the potential is bounded, the global
minimum is always at $\langle y\rangle = \langle V\rangle = \langle W\rangle
= 0,$  so supersymmetry is unbroken if the
potential is bounded.

\subsubsection{$b_1 = 0,\; b_2 = b$.}  This incorporates the Yang-Mills wave
function renormalization at the condensation scale into the K\"ahler potential
for $H$, but leaves the dilaton K\"ahler potential unmodified from its form at
the string scale.  A self-consistent physical interpretation requires
strong coupling: $g^2_c\gg 1$ at the nonperturbative vacuum, where
\beq 2g^2_c = \ell_c = {\ell\over1 + 3b\ell\ln q}> \ell.\eeq
It turns out that that positivity of the kinetic energy together with the
condition $\ell_c>\ell$ requires $\ell_c \le 7.8$. Moreover, if we start in a
region of parameter space where both the potential and the eigenvalues of the
K\"ahler metric are positive, the potential goes to $+\infty$ as we approach
the limiting value of $\ell_c$, which therefore cannot be reached.
Alternatively, the potential is negative for positive eigenvalues of the
K\"ahler metric, and goes to $-\infty$ in the limit.
We conclude that this parameterization is inconsistent with condensation if the
potential is bounded from below.

\subsubsection{No scale case: $b_1 = b_2 = b$.}  This corresponds to replacing
the string coupling constant everywhere by the running coupling constant
evaluated at the condensation scale, and results in a potential that is
automatically positive semi-definite, because it satisfies the general
criterion~\cite{us2} for a no-scale potential.  In this case the potential
is minimized for $W_0 = 0, \;c = W$, so there is no supersymmetry breaking if
$c =0$.   If we take $ F(y) = \alpha y^n\ln(y/\mu)$, we obtain $<W_0>=0$ for
$y= 0$ or $y = \mu$. The interesting case is the latter one ($y\ne0$), for
which
the positivity constraints on the scalar metric are satisfied provided
\beq \mu>1,\qquad \ell_c = {\ell\over 1 + 3b\ell\ln q} = {1\over6b\ln\mu}>
{1\over b} = {1\over .37}. \eeq
Assuming $\alpha,\mu\sim 1,$ and $\ell = g^2/2 = 1/4$, the gravitino squared
mass is given by
\beq m^2_{\tilde G}= {\ell c^2e^G\over (1-x)^3} =
{\alpha^2\mu^{2n}e^{-{1/b\ell_0}}\over9b^2\ell(1-x)^3} \sim{\(5\times 10^{-4}
m_{Pl}\)^2\over
\alpha_c/4\pi} , \eeq
and the compactification radius is
\beq \Lambda_{comp} = \({\rm Re}t{\rm Re}s\)^{-1} \sim
.16 m_{Pl}/(\alpha_c/4\pi)^{1\over 6}, \eeq
where $\alpha_c = \ell_c/2\pi$ is the fine structure constant at the
condensation scale; consistency requires $\alpha_c/4\pi\ge O(1)$.

A pertinent question is whether the use of $S$-duality constraints brings any
qualitatively new features to the problem of gaugino condensation.  We have not
found a satisfactory supersymmetry breaking potential without introducing a
constant $c$, which is the same as the situation without $S$-duality.  A
similar
result was found in~\cite{lalak} where a different definition of $S$-duality
was
used.  Moreover, if we take $M'\to 1$ in (3.6) we find positivity constraints
similar to those in III; taking $n=3$ in the superpotential of III, we recover
the model
studied in~\cite{us,np}, except that the K\"ahler potential has been corrected
to include the Green-Schwarz anomaly cancellation counterterm, with the string
coupling renormalized at the condensation scale $\mu$ ($b_1=b$) so as to
recover
a no-scale effective Lagrangian. Thus the implementation of approximate
invariance under $S$-duality does not seem to qualitatively change the picture
of gaugino condensation, at least in the chiral multiplet formulation.

\subsection{Linear multiplet formulation} There is reason to believe that the
linear multiplet formulation is the correct one for describing
the dilaton supermultiplet from string theory, and, in fact, the Green-Schwarz
counterterm is most easily constructed within this
framework~\cite{counter,linear}.
While this formalism is dual to the chiral formalism in the tree approximation,
it has generally been assumed that this duality may be broken by
nonperturbative
quantum effects.  If this were true, one could interpret the potentials
parameterized in the previous subsection in the following way.  In the absence
of a superpotential, the theory defined by the K\"ahler potential (3.6)
is dual to a theory
containing a linear supermultiplet $L$ and the chiral supermultiplets $T,H$
with K\"ahler potential $K = k(L,Q) + G$, where $k(L,Q)$ is modular
invariant.  The Green-Schwarz counterterm appears~\cite{linear} as a
subtraction constant $V_{GS}$ in the integral equation
\begin{equation}
S+\S + V_{GS}(G,Q) = -\int{dL\over L}{\pp K\over\pp L}.
\end{equation}
obtained from integrating the equation of motion for $L$.  One can choose the
functions $V_{GS}$ and $k$ such that the K\"ahler potential (3.6) is
recovered in
the dual formulation in terms of the chiral supermultiplet $S$. Within this
perspective, one would first perform the duality transformation to cast the
theory in terms of a chiral supermultiplet, and then add a potential induced by
nonperturbative quantum effects which should vanish in the limit of vanishing
gauge coupling constant.  This last requirement coincides with the $S$-duality
constraint imposed in Section 3.A.

However, recent investigations~\cite{lust,burg} suggest that chiral/linear
multiplet duality may not be broken by nonperturbative effects, and that,
in any case, gaugino condensation and the generation of a potential for the
dilaton supermultiplet can be implemented directly within the linear
supermultiplet formulation~\cite{lust,burg,bgt}.  The physical degrees of
freedom of the chiral supermultiplet are the dilaton $\sigma$ and and axion
$a$.
In the classical approximation to effective supergravity theories derived from
string theory, the axion is dual to a three-form $h_{\mu\nu\rho}$ that is the
curl of a two-form potential $b_{\mu\nu}$.  The conventional wisdom has been
that duality is
preserved in the absence of a potential for the axion.  However, it was shown
some time ago~\cite{axion} that interactions for the two-form can be
introduced in such a way that the dual theory contains massive scalars.
In the remainder of this subsection we will consider what the implications of
$S$-duality may be in the general framework of gaugino condensation as
formulated in terms of a linear supermultiplet.  We will see that some of the
difficulties encountered in the chiral multiplet formulation are avoided.

The only subgroup\footnote{The symmetries involving the axion reemerge in the
linear formalism as two independent gauge transformations of the transverse
antisymmetric tensor $b_{\mu\nu}$.}
of \sr~ defined by (2.14) and (2.19) that does not mix the
dilaton $\sigma$ with the axion $a$ is the group of scale transformations:
\beq a^{-1} = d = \lambda,\quad b=c=0, \quad (S' +\S') = \lambda^{-2}
(S + \S),\quad W'_\alpha \to \lambda W_\alpha.\eeq
The transformation law for the linear multiplet $L$ can be inferred from
(3.16), which gives, in the classical limit:
\beq L = (S+\S)^{-1}, \qquad L\to L' = \lambda^2L.\eeq
This symmetry is respected by the modified linearity condition~\cite{linear}
\beq \(\D_{\dot\alpha}\D^{\dot\alpha} - 8R\)L + kW^\alpha W_\alpha = 0, \eeq
where $k$ is a constant, when the Yang-Mills sector is included.

In references~\cite{burg,bgt} a vector supermultiplet $V$ was introduced that
has among its components the components of a linear supermultiplet $L$ and
of a chiral multiplet,
\beq U = -\(\D_{\dot\alpha}\D^{\dot\alpha} - 8R\)V, \eeq
that has the same K\"ahler weight as $W^\alpha W_\alpha$, and moreover
satisfies
the condition:
\beq \(\D^\alpha\D_\alpha - 24R^{\dag}\)U - \(
\D_{\dot\alpha}\D^{\dot\alpha} - 24R\){\bar U} =
{i\over3}\epsilon_{mnpq}\partial^m\Gamma^{npq},\eeq
where $\Gamma_{ABC}$ is a super three-form gauge potential~\cite{form}.
Eq. (3.21) is consistent with the constraint (3.2) if we interpret
$U\sim W^\alpha W_\alpha$ as the condensate chiral multiplet.

In contrast to the case discussed in section 3.A, we do not need to introduce a
``classical'' superpotential for the composite supermultiplet $U$, because the
corresponding term is implicitly included in the kinetic energy term for $V$
just
as, in the linear multiplet formalism for the dilaton coupled to Yang-Mills
fields, the Yang-Mills lagrangian is implicitly included in the lagrangian
for $L$ through the linearity condition (3.19).

In the case of global supersymmetry~\cite{burg,bgt} the ``quantum''
superpotential is the same as in (3.1):
\beq W(U) = {b\over4}U\ln U.\eeq
If no additional operators are introduced in the lagrangian, the complex
scalar field $u$ is a nonpropagating auxiliary field, but a potential is
induced for the dilaton supermultiplet.  The field $u$ can acquire a kinetic
energy term with the introduction of appropriate operators of higher dimension.
This construction was extended to the supergravity case in~\cite{bgt} using
the formalism of superconformal supergravity, with  $N{=}1$
Poincar\'e gauge fixing constraints imposed on the chiral compensator
\cite{KU}.  Here we use the K\"ahler covariant formulation~\cite{bggm}, where
$S$-duality, as defined by (3.18), is more transparent.  In this case the
classical kinetic energy term is just
\beq\L_C = - 3\int d^4\theta E\[\alpha + {1\over3}VV_{GS}\] , \eeq
where $\alpha$ is a constant, and $V_{GS} = -bG$ is the Green-Schwarz
counterterm introduced in (3.16).
The nonperturbatively induced superpotential term is
\beq \L_Q = {b\over8}\int d^4\theta{E\over R} U\ln(e^{-K/2}U) + {\rm h.c.},
\qquad K = \ln V + G ,\eeq where the argument of the log, which must be a
superfield of K\"ahler chiral weight $w=0$, can be understood~\cite{np} in
terms of the ratio
of the infrared cut-off ($U^{1\over3}$) and the effective ultraviolet cut-off
($e^{K/6}$). Under the modular transformation (3.5), with
\beq V'(\theta) = V(\theta'), \qquad U'(\theta) = e^{{1\over2}(\F - F)}
U(\theta') ,\eeq
we have
\beq \delta\L_C = - \delta\L_Q = b\int d^4\theta EV(F + \F) =
{b\over8}\int d^4\theta {E\over R}UF + {\rm h.c.}.\eeq

If the kinetic term for the condensate is dominated by field theory
quantum corrections and/or string corrections analogous to (3.10), that arise
only from the dilaton/Yang-Mills/gravity sector interactions, then $S$-duality
can be used as a guide to their construction.  For example, in the general
formalism described in~\cite{bggm}, terms in the locally supersymmetric
lagrangian are of the generic form
\beq \L = {1\over2}\int {E\over R}\Phi + {\rm h.c.}, \eeq
where $\Phi$ is a chiral supermultiplet ($\D^{\dot\alpha}\Phi = 0$) with
K\"ahler chiral weight $ w(\Phi) = + 2$. A special case is
\beq \L = \int E \phi, \;\;\;\; w(\phi) = 0, \;\;\;\;
\Phi = -{1\over8}\(\D_{\dot\alpha}\D^{\dot\alpha} - 8R\)\phi. \eeq
Superfields $\Phi,\phi$ can be constructed from $S$-duality invariant forms
such as
$$ V^{-n}\D^m V^n, \;\;\;\; \D^m\ln V, \;\;\;\; \D = \D^\alpha
,\D_{\dot\alpha}.
$$
Terms similar to these were used in~\cite{bgt} to generate effective
lagrangians for a dynamical condensate.  The restricted class of models studied
(which do not include a constant $c$ in the superpotential) do not break
supersymmetry at the vacuum.

\section{Conclusions}

We have explicated the gauge and dilaton superfield transformations under
the \sr ~group of duality rotations that includes weak/strong coupling duality,
$S\to 1/S$, as a group element.  \sr ~is a symmetry of the
Yang-Mills/dilaton/gravity sector in
effective Lagrangians obtained from superstring theory, and we have studied its
implications for models with a dynamical condensate.  In the linear multiplet
formulation for the dilaton, the only remnant of \sr ~is a scale transformation
that can nevertheless be used to constrain the operators appearing in the
lagrangian, and in fact this formulation appears to be the more natural
framework
for describing a composite superfield condensate.  None of the models studied
so
far have produced a bounded potential with supersymmetry breaking at the vacuum
without the introduction of a constant term in the superpotential.

\vskip .2in
\noindent{\bf Acknowledgements.} We thank Bruno Zumino for discussions, and
Kamran Saririan and Yi-Yen Wu for a careful reading of the manuscript.  This
work was supported in part by the
Director, Office of Energy Research, Office of High Energy and Nuclear Physics,
Division of High Energy Physics of the U.S. Department of Energy under Contract
DE-AC03-76SF00098 and in part by the National Science Foundation under grant
PHY-90-21139.

\end{document}